# Uplinks Analysis and Optimization of Hybrid Vehicular Networks

**Shikuan Li[1], Zipeng Li[1], Xiaohu Ge[1], Yonghui Li[2]**
[1] School of Electronic Information and Communications, Huazhong University of Science and Technology
Wuhan, Hubei 430074 - China
[e-mail: li_shikuan, xhge, zipengli91@mail.hust.edu.cn]
[2] School of Electrical and Information Engineering
The University of Sydney, Sydney
[e-mail: yonghui.li@sydney.edu.au]
*Corresponding author: Xiaohu Ge

***Abstract***

5G vehicular communication is one of key enablers in next generation intelligent transportation system (ITS), that require ultra-reliable and low latency communication (URLLC). To meet this requirement, a new hybrid vehicular network structure which supports both centralized network structure and distributed structure is proposed in this paper. Based on the proposed network structure, a new vehicular network utility model considering the latency and reliability in vehicular networks is developed based on Euclidean norm theory. Building on the Pareto improvement theory in economics, a vehicular network uplink optimization algorithm is proposed to optimize the uplink utility of vehicles on the roads. Simulation results show that the proposed scheme can significantly improve the uplink vehicular network utility in vehicular networks to meet the URLLC requirements.

# 1. Introduction

**I**n recent years, the intelligent transportation system has become a research hotspot which is one of the most promising application scenarios in the fifth generation (5G) mobile communication systems [1-2]. Different from the traditional mobile communication networks, the critical information generated by the vehicles need to be transmitted with very low latency and extremely high reliability [3-4]. It is difficult for traditional vehicular networks to realize ultra-reliable and low latency communication (URLLC) [5]. How to design a dynamical vehicular network to achieve URLLC presents a great challenge.

Currently, there are mainly two types of network structures used to realize vehicular networks in the existing studies, a centralized network structure based on cellular networks [6-8] and a distributed vehicular network structure, also known as the vehicular ad-hoc network (VANET) [9-11]. The advantage of LTE based centralized vehicular networks was discussed in [6]. In [7], a TD-LTE based vehicular network structure was introduced and the transmission reliability under different transmission distances was studied. The analysis results showed that LTE is a promising technology to support the long range vehicle to infrastructure (V2I) communications. However, the high latency remains as a key issue in LTE based vehicular networks. Since the low transmission reliability of messages in vehicular networks lead to the large resend time which does not guarantee the short handling time for security messages when there is a large distance between vehicle and road side unit. In fact, as one of the most important performance criteria in vehicular networks, the latency of transmitting critical information, such as the safety-related messages is one of key metrics to evaluate performance of a vehicular network [8].

In distributed VANET, vehicles on the road can communicate with each other via vehicle to vehicle (V2V) links and vehicle clusters are formed to provide a flexible network structure. In [9], an access reliability model in VANET was proposed to analyze the vehicular access reliability with a fixed latency constraint. Connectivity probability under different traffic density for V2V communication scenarios in one- and two-way platoon-based VANETs was investigated in [10]. In [11], a one-dimensional multi-hop broadcasting model considering the different safe distances between vehicles was proposed to analyze the broadcasting success probability and multi-hop latency in VANETs.

However, either centralized or distributed vehicular network structure cannot meet the URLLC requirements of future vehicular networks and can only satisfy parts of the vehicular network requirements [12]. To tackle this problem, in this paper, a hybrid vehicular network model which can adjust network structure dynamically is proposed to achieve URLLC, as a Pareto improvement strategy is deployed in the hybrid vehicular network which can lead to an optimal vehicular network utility under different distances between vehicles and the associated RSU. The contributions of this paper are summarized as follows.

1) To analyze the latency and reliability of messages in vehicular networks, a novel vehicular network utility model based on Euclidean norm theory is proposed for the hybrid vehicular networks.
2) By using the Pareto optimization, an uplink optimization algorithm is proposed to improve the vehicular network utility of uplinks in the hybrid vehicular networks.
3) Simulation results show that the proposed uplink optimization algorithm can significantly improve the uplink reliability and reduce the uplink latency. The vehicular network utility of the Pareto improvement strategy is improved up to 15% compared with random relay strategy and cellular access strategy in hybrid vehicular networks.

## 2. System Model

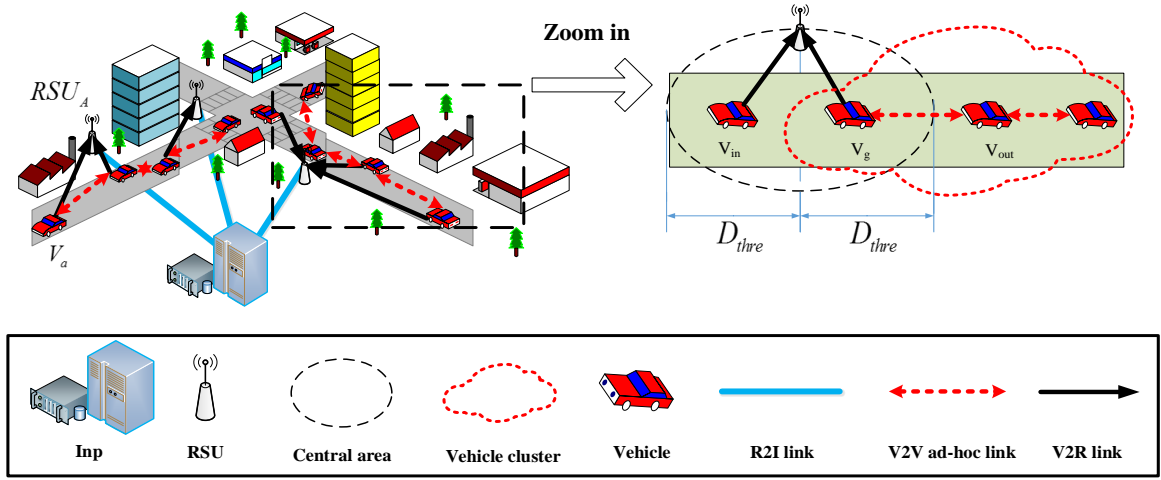

**Fig. 1**. System model

We consider typical urban vehicular network shown in **Fig. 1**. In this network, road side units (RSUs) are distributed uniformly along the roads. These RSUs can provide vehicular network access services for vehicles through the vehicle to RSU (V2R) wireless uplinks. Due to the limited computation capability of RSUs, in most application scenarios, RSUs are only responsible for the message forwarding, and the complex calculation service is hard to be done in RSUs. Therefore, an infrastructure provider (Inp) is deployed to gather information from RSUs and provide calculation services for vehicle applications. Messages or requests generated by vehicles are first transmitted to RSUs and these messages and requests will be forwarded to Inp through RSU to Inp (R2I) links.

In this paper we propose a hybrid vehicular network framework, integrating both centralized and distributed vehicular network structures. In the centralized structure, vehicles directly access to RSUs. Vehicles on the road can directly communicate with RSU through V2R links. In the distributed vehicular network structure adjacent vehicles communicate with each other via vehicle to vehicle (V2V) links and form a cluster. Messages can be stored and forwarded in each relay vehicle. The vehicle with the best link to RSU in the cluster is selected as the gateway vehicle. Any vehicles in the cluster can transmit messages to the gateway vehicle via multi-hop V2V links. The gateway vehicle forwards this messages to the associated RSU.

Compared with distributed vehicular network structure, the centralized vehicular network structure can effectively reduce the transmission latency of messages when the transmission distance is short because there is no unnecessary store and forward latency. However, the link reliability of the centralized vehicular network structure depends on the distance between the vehicle and the associated RSU. When the distance between the vehicle and associated RSU is very long, the link reliability between the vehicle and RSU becomes very low. In this case, the distributed vehicular network structure can be adopted to guarantee the link reliability via multi-hop V2V links. Since the distance between adjacent vehicles is shorter than or equal to the distance between the source vehicle and the associated RSU in multi-hop links, the reliability of the multi-hop links can be significantly improved. However, such a multi-hop transmission introduces a long delay. Therefore, the total latency of the multi-hop V2V links becomes very high when the number of relay vehicles is large.

To meet different requirements of vehicular networks, in this paper a hybrid vehicular network framework is proposed by effectively integrating these two type of vehicular network structures as shown in the right side of **Fig. 1**. Since the Inp can gather information from RSUs and vehicles, the Inp is served as the central controller of this framework and has access to the following information: 1) locations of RSUs and vehicles; 2) state information of V2R and V2V links. A distance threshold $D_{thre}$ based on which the controller determines which type of network structure is selected for the hybrid vehicular network framework. When the distance between the vehicle and associated RSU is less than or equal to $D_{thre}$, the centralized vehicular network structure is adopted for messages transmission. In this case, the vehicle $V_{in}$ communicates with the associated RSU directly. When the distance between the vehicle and associated RSU is larger than $D_{thre}$, the distributed vehicular network structure is adopted for messages transmission. In this case, the vehicle $V_{out}$ transmits messages via V2V ad-hoc links to the gateway vehicle $V_g$ which is closest to the associated RSU in the vehicle cluster. The gateway vehicle $V_g$ forwards these messages to the associated RSU.

## 3. Vehicular Network Utility Function

To evaluate the performance of hybrid vehicular network framework, a unified vehicular network utility function $\Omega$ is defined as

$$\Omega \stackrel{def}{=} \{w(x_i) + \varphi(x_i)\} 1\{x_i \in \mathbb{C}\}, \tag{1}$$

where $w(x_i)$ is the latency utility function, $\varphi(x_i)$ is the reliability utility function, $1\{\ \}$ is an indicator function, $x_i$ is the distance between the vehicle $V_a$ and the associated RSU, and the symbol $\mathbb{C}$ is the set of all possible distances between the vehicle $V_a$ and the associated RSU. In (1) the vehicular network utility function $\Omega$ is defined as an additive functional form of the deterministic effect of the latency utility function and reliability function. By utilizing the Euclidean norm with weights, the effect of latency and reliability on vehicular network performance can be evaluated in the vehicular network utility function space. As mentioned in Section II, a distance threshold $D_{thre}$ is used for the controller to determine which type of network structures is selected for the hybrid vehicular network framework. Therefore, the vehicular network utility function with the centralized or distributed vehicular network structure can be classified according to the distance between the vehicle $V_a$ and the associated RSU $x_i$.

When the distance between the vehicle $V_a$ and the associated RSU $x_i$ is less than the distance threshold $D_{thre}$, high reliability of links can be ensured. In this case, the vehicular network utility depends on the latency utility. Since there is no store-and-forward latency in centralized vehicular network structure compared with distributed vehicular network structure, centralized vehicular network structure is adopted in this case. The vehicular network utility function with centralized vehicular network structure $\Omega_C$ can be expressed as

$$\Omega_C \stackrel{def}{=} \{w^c(x_i) + \varphi^c(x_i)\} 1\{x_i \in \mathbb{C}, x_i \leq D_{thre}\}$$
$$\stackrel{a}{=} \left| (\alpha_c w^c(\mathbb{L}))^2 + (\beta_c \varphi^c(\mathbb{L}))^2 \right|^{1/2}, \qquad (2)$$

where the step $a$ is the Euclidean norm operation, $\mathbb{L}$ denotes the wireless uplink from the vehicle to the associated RSU, $w^c(\mathbb{L})$ and $\varphi^c(\mathbb{L})$ are the latency utility function and reliability utility function in centralized vehicular network structure, respectively. $\alpha_c$ is the weight factor of latency utility which indicates the weight of latency utility in the vehicular network utility function. Similarly, $\beta_c$ is the weight factor of reliability utility.

The uplink in the hybrid vehicular network is configured as a wireless link between the vehicle and the Inp. Messages of applications are produced in vehicles and need to be processed in Inp, so a wireless link is configured to upload these messages. Considering the URLLC in vehicular networks, these messages need to be transmitted with a short latency and high success probability. Therefore, the uplink utility function is introduced to evaluate the network performance under different uplinks with different transmission distances. When the distance between the vehicle $V_a$ and associated RSU is larger than the distance threshold $D_{thre}$, to overcome the negative impact of long distance link on the reliability of vehicular network, the distributed vehicular network structure is adopted. The vehicular network utility function under distributed vehicular network structure $\Omega_d$ is expressed as

$$\Omega_d \stackrel{def}{=} \{w^d(x_i) + \varphi^d(x_i)\} 1\{x_i \in \mathbb{C}, x_i > D_{thre}\}$$
$$\stackrel{b}{=} \left| (\alpha_d w^d(\mathbb{L}))^2 + (\beta_d \varphi^d(\mathbb{L}))^2 \right|^{1/2}, \qquad (3)$$

where $w^d(\mathbb{L})$ and $\varphi^d(\mathbb{L})$ represent the latency utility function and the reliability utility function under distributed vehicular network structure, respectively. The weight factor of latency utility and reliability is defined as $\alpha_d$ and $\beta_d$.

To simplify the notation of the latency and reliability utility function, we set $\kappa \in \{c, d\}$ and then the vehicular network latency utility function $w^\kappa(\mathbb{L})$ and reliability utility function $\varphi^\kappa(\mathbb{L})$ is extended as

$$w^\kappa(\mathbb{L}) = e^{\left(\frac{T_{req} - T(\mathbb{L})}{T_{req}}\right)}, \qquad (4)$$

$$\varphi^\kappa(\mathbb{L}) = e^{\left(\frac{P(\mathbb{L}) - P_{req}}{P_{req}}\right)}, \qquad (5)$$

where $T(\mathbb{L})$ is the uplink transmission latency when the uplink is $\mathbb{L}$, $P(\mathbb{L})$ is the probability of successful message transmission, i.e., the transmission reliability. The latency and reliability requirements are denoted as $T_{req}$ and $P_{req}$, respectively. Based on (4) and (5), the values of these two utility functions are always larger than 0. When the latency and reliability requirements are both satisfied in vehicular networks, i.e., $T(\mathbb{L}) \leqslant T_{req}$ and $P(\mathbb{L}) \geqslant P_{req}$, the values of latency and reliability utility functions are always large than or equal to 1. Otherwise, the values of latency and reliability utility functions are always large than 0 and less than 1 if the latency and reliability requirements are both not satisfied, i.e., $T(\mathbb{L}) > T_{req}$ and $P(\mathbb{L}) < P_{req}$.

## 4. Uplinks Analysis of Hybrid Vehicular Networks

In this section the uplinks of hybrid vehicular networks, i.e., the wireless link from the vehicle to the associated RSU and the wired link from the RSU to the Inp are investigated.

### 3.1 Uplink Latency in Hybrid Vehicular Networks

In this case, the whole uplink latency in hybrid vehicular networks $T_{V-I}$ consists of two parts: the first part comes from wireless V2R uplinks, denoted as the wireless transmission latency $T_{V-R}$; the second part is the transmission latency in wired links between RSUs and Inp, denoted as $T_{R-I}$. Therefore, the uplinks latency $T_{V-I}$ in hybrid vehicular networks is expressed as

$$T_{V-I} = T_{V-R} + T_{R-I}. \tag{6}$$

Assume that millimeter wave transmissions are used for V2V and V2R communications. In detail, 72 GHz millimeter wave path loss model is used to analyze the path loss of wireless signal transmissions [13-14]. Based on the results in [14], when the distance between the transmission node and receiving node is $d$, the path loss is expressed as

$$PL[dB](d) = 69.6 + 20.9\log(d) + \xi, \xi \sim \mathcal{N}(0, \sigma^2), \tag{7}$$

where $\xi$ is the shadow fading coefficient, and $\sigma$ is the standard deviation of shadow fading. Since the beam forming technology is widely adopted in millimeter wave transmissions, the urban scenario in this paper is regarded as a noise-limited scenario [15], the signal to noise ratio (SNR) in the receiving node is derived as

$$SNR[dB] = P_{tx}[dB] - PL[dB] - N_0 W_{mmWave}[dB], \tag{8}$$

where $P_{tx}$ is the transmission power of vehicles, $N_0$ is the Gaussian white noise power spectral density and $W_{mmWave}$ is the bandwidth of millimeter wave. When the SNR at the receiving node is larger than the SNR threshold $\theta$, the messages can be received successfully. Therefore, the transmission success probability in the uplink with the single hop $P_{hop}$ is derived as

$$\begin{aligned} P_{hop} &= P\big(SNR[dB] \geq \theta[dB]\big) \\ &= P\big(P_{tx}[dB] - PL[dB] - N_0 W_{mmWave}[dB] \geq \theta[dB]\big) \\ &= P\big(PL[dB] \leq P_{tx}[dB] - \theta[dB] - N_0 W_{mmWave}[dB]\big) \\ &= P\big(\xi \leq P_{tx}[dB] - \theta[dB] - N_0 W_{mmWave}[dB] - 69.6 - 20.9 \log_{10} d\big) \\ &= \frac{1}{2}\left(1 + erf\left(\frac{\psi(d)}{\sqrt{2}\sigma}\right)\right) \end{aligned} \quad , \quad (9a)$$

$$\psi(d) = P_{tx}[dB] - \theta[dB] - N_0 W_{mmWave}[dB] - 69.6 - 20.9 \log_{10} d, \quad (9b)$$

where $erf()$ is the Gaussian error function. Assumed that messages are encapsulated in packets for wireless transmissions in hybrid vehicular networks. The transmitting slot of a packet is denoted as $t_{slot}$. Therefore, the transmission latency $T_{hop}$ in the single hop is expressed as

$$T_{hop}(d) = t_{slot}\big/P_{hop} = \frac{2 t_{slot}}{1 + erf\left(\frac{\psi(d)}{\sqrt{2}\sigma}\right)}. \quad (10)$$

When centralized vehicular network structure is adopted in hybrid vehicular networks, V2R uplink $\mathbb{L}_c$ is one hop wireless link. Assumed the distance of uplink $\mathbb{L}_c$ is $d_c$. The wireless transmission latency under centralized vehicular network structure $T_{V-R-Centralized}(\mathbb{L}_c)$ is expressed as

$$T_{V-R-Centralized}(\mathbb{L}_c) = T_{hop}(d_c) = \frac{2 t_{slot}}{1 + erf\left(\frac{\psi(d_c)}{\sqrt{2}\sigma}\right)}. \quad (11)$$

When distributed vehicular network structure is adopted in hybrid vehicular networks, V2R uplink $\mathbb{L}_d$ is the multi-hop wireless link. In this case, the multi-hop wireless uplink $\mathbb{L}_d$ is expressed as

$$\mathbb{L}_d = \{D_1, D_2, \cdots, D_y\}\big(y = 1, 2, \cdots, N_{hop}\big), \quad (12)$$

where $D_y$ is the $y-th$ hop of multi-hop uplink and $N_{hop}$ denotes the number of the hops. Let $t_{proc}$ denotes the store-and-forward latency in each relay vehicle and $d_y$ denotes the distance of $D_y$. The wireless transmission latency under distributed vehicular network structure $T_{V-R-Distributed}(\mathbb{L}_d)$ is derived as

$$T_{V-R-Distributed}(\mathbb{L}_d) = \sum_{D_y \in \mathbb{L}_d} T_{hop}(d_y) + (N_{hop}-1)t_{proc}. \tag{13}$$

In this paper RSUs are only responsible for the transmission of messages and Inps provide data storage and calculation services. Different from the wireless transmission latency, empirical measurements show that there is a long-tail effect on the latency of data packets in wired networks [16]. Therefore, exponential distribution models are widely used to analyze the latency of data packet in wired links. Without loss of generality, R2I wired uplink latency $T_{V-I}$ is expressed as [16]

$$T_{R-I} = \beta_w \frac{\rho_R}{\rho_I} \int_0^\infty 2\rho_I \pi r^2 e^{-\pi \rho_I r^2} dr = \frac{1}{2}\beta_w \rho_R \rho_I^{-3/2}, \tag{14}$$

where $r$ is the intermediate variable, $\rho_R$ is the density of RSU, $\rho_I$ is the density of Inp, $\beta_w$ is a scaling factor that indicates the wired channel state.

Assume that the distance between two adjacent RSUs is $L$ and the number of RSUs associated with a Inp is $N_{RSU}$. Therefore, the density of RSU in this area is expressed as

$$\rho_R = \rho_I N_{RSU}. \tag{15}$$

Let $\mathbb{R}_{Inp}$ denote a typical vehicular network area which includes a Inp and $N_{RSU}$ RSUs. Based on the results in [17-18], the PDF of the area of $\mathbb{R}_{Inp}$ is expressed as

$$f_{\mathbb{R}_{Inp}}(\varepsilon) = \frac{(b\rho_I)^a}{\Gamma(a)} \varepsilon^{a-1} e^{-b\varepsilon \rho_I}, \tag{16}$$

where $\Gamma(\varepsilon) = \int_0^\infty \tau^{\varepsilon-1} e^{-t} d\tau$ is the Gamma function, $\varepsilon$ and $\tau$ is the intermediate variable, $a$ is the shape parameter and $b\rho_I$ is the inverse scale parameter for a Gamma distribution [19-20]. The road length in the coverage of Inp is derived as

$$L_{\mathbb{R}_{Inp}}(\varepsilon) = \int_0^\infty \rho_{road} \frac{(b\rho_I)^a}{\Gamma(a)} \varepsilon^{a-1} e^{-b\varepsilon \rho_I} d\varepsilon, \tag{17}$$

where $\rho_{road}$ is the road density in urban environments. Assume that RSUs are governed by a uniform distribution. The expectation of $N_{RSU}$ is derived as

$$\mathrm{E}(N_{RSU}) = \frac{1}{L}\int_0^\infty \rho_{road} \frac{(b\rho_I)^a}{\Gamma(a)} \varepsilon^{a-1} e^{-b\varepsilon\rho_I} d\varepsilon. \tag{18}$$

According to (14), (15) and (18), the R2I wired uplink latency $T_{R-I}$ in hybrid vehicular network is derived as

$$\begin{aligned} T_{R-I} &= \beta_w \frac{\rho_R}{\rho_I} \int_0^\infty 2\rho_I \pi r^2 e^{-\pi\rho_I r^2} dr \\ &= \beta_w \rho_R \rho_I^{-3/2} \\ &= \frac{1}{2} N_{RSU} \beta_w \rho_I^{-1/2} \\ &= \frac{1}{2L} \beta_w \rho_I^{-1/2} \int_0^\infty \rho_{road} \frac{(b\rho_I)^a}{\Gamma(a)} \varepsilon^{a-1} e^{-b\varepsilon\rho_I} d\varepsilon \end{aligned} \tag{19}$$

### 3.2 Reliability of Uplinks in Hybrid Vehicular Networks

Since the link outage problem in hybrid vehicular networks is mainly caused by the wireless links, the probability of successful message transmission in wireless links $P(\mathbb{L})$ is used to represents the reliability of hybrid vehicular networks.

When centralized vehicular network structure is adopted in hybrid vehicular networks, the uplink success probability $P(\mathbb{L}_c)$ equals to the one-hop transmission probability and is expressed as

$$P(\mathbb{L}_c) = P_{hop}(d_c) = \frac{1}{2}\left(1 + erf\left(\frac{\psi(d_c)}{\sqrt{2}\sigma}\right)\right). \tag{20}$$

When distributed vehicular network structure is adopted in hybrid vehicular networks, the uplink success probability of the uplinks $P(\mathbb{L}_d)$ equals to the product of success probability of each single hop and is expressed as

$$P(\mathbb{L}_d) = \prod_{D_y \in \mathbb{L}_d} P_{hop}(d_y). \tag{21}$$

## 5. Uplinks Optimization of Hybrid Vehicular Networks

Let $\Lambda$ denotes the set of all possible uplinks for vehicle $V_a$ in hybrid vehicular networks. Obviously, set $\Lambda$ contains not only the uplinks of centralized vehicular network structure but also the uplinks of distributed vehicular network structure. To achieve a high network utility, the vehicle $V_a$ needs to choose a suitable uplink with low latency and high reliability from $\Lambda$. Compared with multi-hop links in distributed vehicular network structure, the store-and-forward latency can be avoided in centralized vehicular network structure. However, the reliability of distributed vehicular network structure can be improved by shorting the wireless transmission distance compared to the centralized vehicular network structure. Therefore, there is a tradeoff between latency and reliability for optimizing uplinks in the hybrid vehicular networks.

In order to maximize $\Omega(\mathbb{L})$, a model called Pareto improvement which is widely used in field of economics is introduced to optimize the selection of uplinks from set $\Lambda$. In the field of economics, the concept of Pareto improvement is to make any individual or preference criterion better without making at least one individual or preference criterion worse [21]. In this case, the uplink latency utility and reliability utility can be treated as two individuals in Pareto improvement model and the optimization objective function is $\Omega(\mathbb{L})$ in hybrid vehicular networks. Based on the Pareto improvement, a new uplink $\mathbb{L}'$ in set $\Lambda$ will be selected if $\mathbb{L}'$ leads to a shorter latency without reducing the reliability or leads to a higher reliability without increasing the latency compared with the current link $\mathbb{L}$, i.e., $T(\mathbb{L}') < T(\mathbb{L}) \& P(\mathbb{L}') \geqslant P(\mathbb{L})$ or $P(\mathbb{L}') > P(\mathbb{L}) \& T(\mathbb{L}') \leqslant T(\mathbb{L})$.

Therefore, the optimization problem is to find an optimal uplink $\mathbb{L}_{opt}$ which maximizes the network utility based on Pareto improvement. The optimization problem is formulated as

$$\underset{\mathbb{L}_{opt} \in \Lambda}{\operatorname{argmax}} \ \Omega(\mathbb{L}_{opt})$$

*subject to*

$$(e1) \ \sum_{D_y \in \mathbb{L}_{opt}} D_y \geq D_0 \qquad\qquad . \qquad (22)$$

$$(e2) \ \{P(\mathbb{L}_{opt}) \geq P_{req}\} \& \{T(\mathbb{L}_{opt}) \leq T_{req}\}$$

$$(e3) \ \{\forall \mathbb{L} \in \Lambda\} \& \{\mathbb{L} \neq \mathbb{L}_{opt}\}, \{T(\mathbb{L}_{opt}) \leq T(\mathbb{L}) \& P(\mathbb{L}_{opt}) \geq P(\mathbb{L})\}$$

where the constraint $(e1)$ indicates that if the uplink $\mathbb{L}$ is a multi-hop link, the sum of distances in all hops should be larger than or equals to $D_0$ which is the distance between the vehicle $V_a$ and RSU in the direct V2R link. The constraint $(e2)$ guarantees maximum latency and minimum reliability requirements that can be tolerated in vehicular networks. The constraint $(e3)$ is the condition of Pareto improvement.

In order to solve the optimization problem (22), an iterative algorithm, i.e., the uplink pareto improvement algorithm for hybrid vehicular network is developed as follows.

**Algorithm 1**: Uplink Pareto Improvement Algorithm for Hybrid Vehicular Network

Begin

1: Initialize $\Omega(\mathbb{L}_{opt})$, $D_0$, $T_{req}$, $P_{req}$, $T(\mathbb{L}_{opt})$, $P(\mathbb{L}_{opt})$;

2: Get number of vehicles between transmission vehicle and it's closest RSU as $N_v$, get all vehicles position as $\chi_j (j=1,2,\cdots,N_v)$;

3: If $N_v = 0$

4:     $T(\mathbb{L}_{opt}) = T_{hop}(D_0)$;

5:     $P(\mathbb{L}_{opt}) = P_{hop}(D_0)$;

6: else

7:     For $n = 1:1:N_v$ do

8:       For $k = 1:1:n$

9:         Initialize $T(\mathbb{L})$, $P(\mathbb{L})$, $\Omega(\mathbb{L})$;

10:         Set target positions $\lambda_k = \dfrac{D_0}{n+1} \cdot k$;

11:         For $i = 1:1:N_v$

12:           Find the closest vehicle $V_i$ as the kth relay vehicle which satisfy $\forall \chi_j (j=1,2,\cdots,N_v), abs(\chi_j - \lambda_k) \leq abs(\chi_j - \lambda_k)$;

13:           $\lambda_k = \chi_j$;

14:         End for

15:         $T(\mathbb{L}) = T(\mathbb{L}) + T_{hop}(\lambda_k - \lambda_{k-1})$;

16:         $P(\mathbb{L}) = P(\mathbb{L}) \cdot P_{hop}(\lambda_k - \lambda_{k-1})$;

17:         Get $\Omega(\mathbb{L})$ by equation (3);

18:       End For

19:       $\mathbb{L} = \{\lambda_1, \lambda_2 - \lambda_1, \cdots, \lambda_k - \lambda_{k-1}\}$

20:       If $(\Omega(\mathbb{L}) \geq \Omega(\mathbb{L}_{opt})) \& (T(\mathbb{L}) \leq T(\mathbb{L}_{opt}) \& P(\mathbb{L}) \geq P(\mathbb{L}_{opt}))$

21:         $\mathbb{L}_{opt} = \mathbb{L}$;

| | |
|---|---|
| 22: | $\Omega(\mathbb{L}_{opt}) = \Omega(\mathbb{L})$; |
| 23: | End if |
| 24: | End For |
| 25: | End if |

The Pareto Optimality link is $\mathbb{L}_{opt}$ with the best $\Omega(\mathbb{L}_{opt})$.

## 6. Numerical Results and Discussion

To evaluate the performance of the proposed algorithm, the configuration parameters for the numerical simulations are summarized in **Table 1**.

Table 1. configuration parameters

| Parameter | Value |
|---|---|
| The density of Inp $\rho_I$ | $10^{-7} \sim 3 \times 10^{-7}$ per square meter[15-16] |
| The vehicle density on the road $\rho_v$ | 0.08~0.24 vehicle per meter [22] |
| The transmission power of vehicles $P_{tx}$ | 30 dBm |
| The noise power density $N_0$ | -174 dBm/Hz [14, 24] |
| The standard deviation of shadow fading $\sigma$ | 5 |
| The duration of a slot $t_{slot}$ | 50 microseconds [15] |
| The SNR threshold $\theta$ | 5 dB |
| The scaling factor of wired links between RSU and Inp $\beta_w$ | $5 \times 10^{-4}$ [16] |
| The road density under urban environment $\rho_{road}$ | 0.004 meter per square meter [23] |
| $a$ | 3.61 [19] |
| $b$ | 3.57 [19] |
| The uplink latency requirement $T_{req}$ | 1 millisecond [11] |
| The uplink reliability requirement $P_{req}$ | 0.9 |
| $\alpha_c, \alpha_d, \beta_c, \beta_d$ | 0.5 |

**Fig. 2** shows the uplink success probability P with respect to the distance $D_0$ between vehicle and the associated RSU considering different vehicular network structures and vehicle densities $\rho_V$ on the road. When the vehicular network structure is fixed, it is shown that uplink success probability decreases as the between vehicle $V_a$ and the associated RSU increases. When the distance between vehicle and the associated RSU is fixed, the uplink success probability of centralized vehicular network structure is less than the uplink success probability of distributed vehicular network structures. Furthermore, when the distance between vehicle and the associated RSU is fixed, the uplink success probability of distributed vehicular network structure increases with the vehicle density on the road.

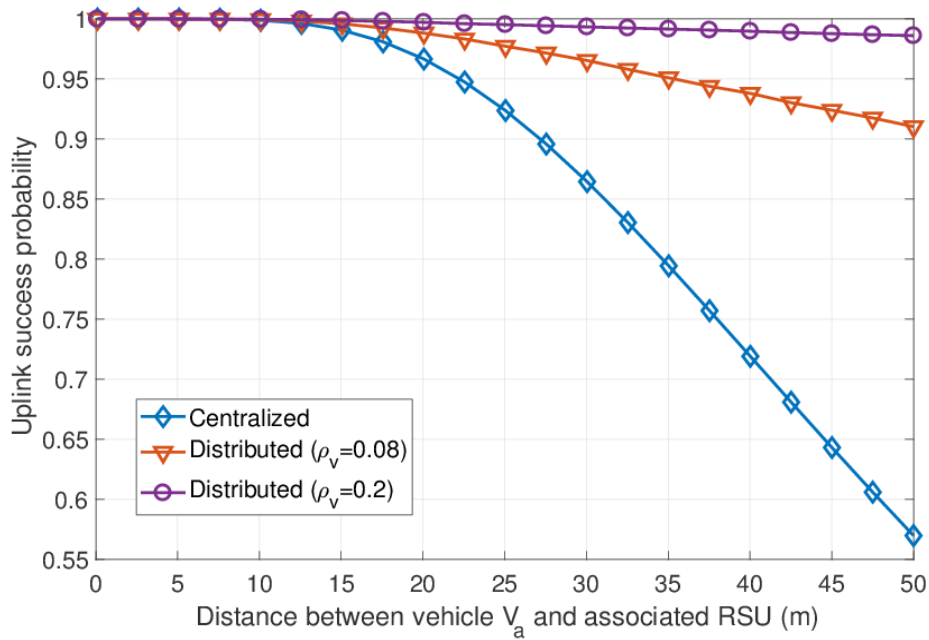

**Fig. 2**. Uplink success probability with respect to the distance between vehicle $V_a$ and the associated RSU

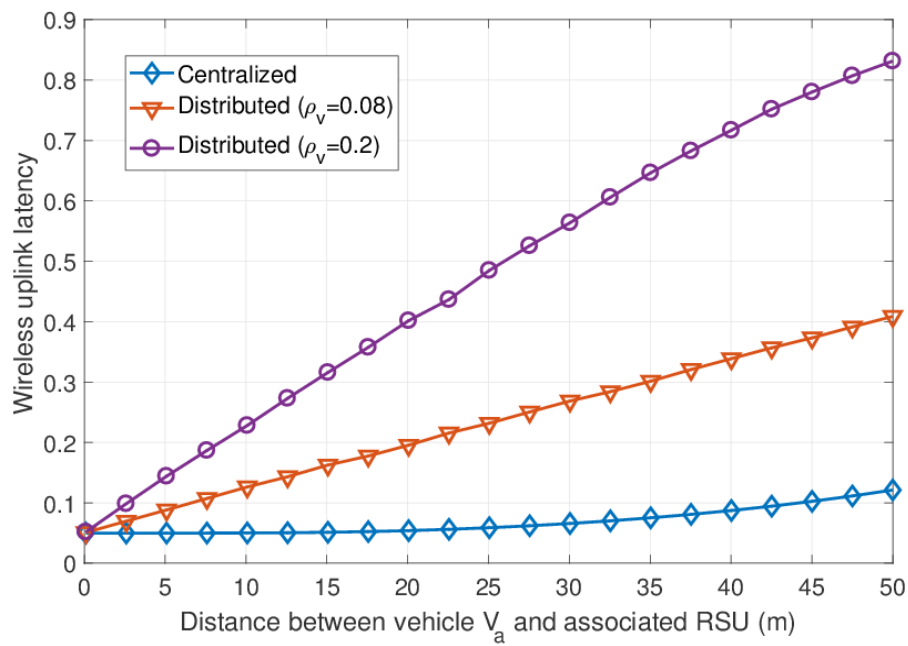

**Fig. 3**. Wireless uplink V2R latency (ms) with respect to the distance between vehicle $V_a$ and the

Considering centralized vehicular network structure and distributed vehicular network structure, the impact of distance between vehicle $V_a$ and the associated RSU on the wireless uplink V2R latency $T_{V-R}$ is investigated in **Fig. 3**. When the vehicular network structure is fixed, the wireless uplink V2R latency increases with the distance between vehicle $V_a$ and the associated RSU. When the distance between vehicle $V_a$ and the associated RSU is fixed, the wireless uplink V2R latency of centralized vehicular network structure is less than the wireless uplink V2R latency of distributed vehicular network structure.

**Fig. 4** illustrates the wired uplink R2I latency $T_{R-I}$ with respect to the Inp density considering different distances between adjacent RSUs. When the distance between adjacent RSUs is fixed, the wired uplink R2I latency decreases with Inp density. When the Inp density is fixed, the wired uplink R2I latency decreases as distance L between adjacent RSUs increases. When the Inp density is fixed, the expectation of distance between Inp and vehicle is fixed. When the distance L between adjacent RSUs is increased, the distance between vehicles and the associated RSU is decreased. As a consequence, the wired uplink R2I latency is increased.

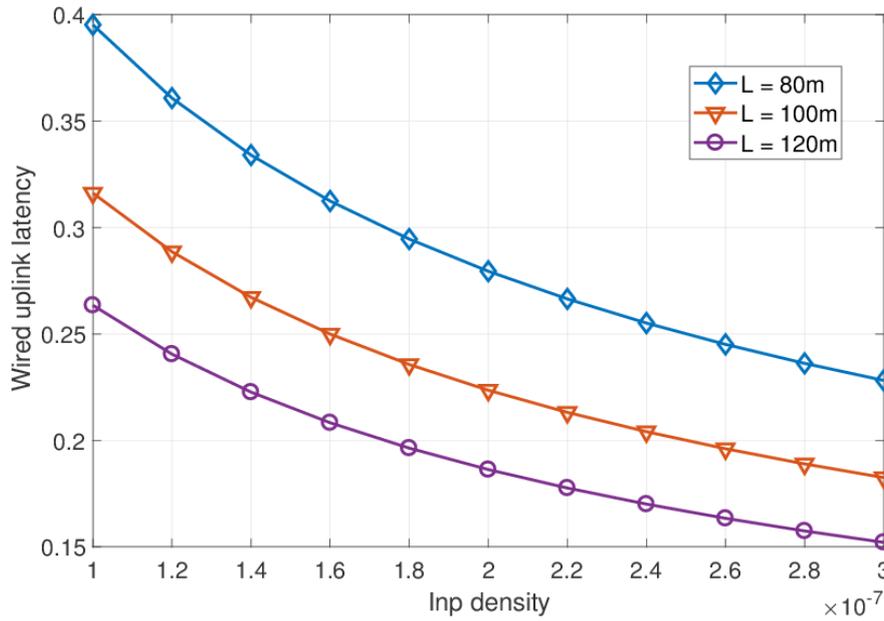

**Fig. 4**. Wired uplink R2I latency (ms) with respect to the Inp density and distance between adjacent RSUs

**Fig. 5** shows the vehicular network utility with respect to the distance between vehicle and the associated RSU considering different vehicular network structures. When the vehicular network structure is fixed, the vehicular network utility is a decreasing function of distance between vehicle and RSU. When the distance between vehicle $V_a$ and the associated RSU is less than 27 meters, the vehicular network utility of centralized vehicular network structure is larger than that of distributed vehicular network structure. This indicates that vehicular network utility is mainly affected by the latency when the distance between vehicle $V_a$ and the associated RSU is short, i.e., the distance between vehicle $V_a$ and the associated RSU is less than 27 meters. When the distance between vehicle $V_a$ and the associated RSU is larger than or equal to 28.3 meters, the vehicular network utility of centralized vehicular network structure is less than that of distributed vehicular network structure. This result implies that vehicular network utility is mainly affected by reliability when the distance between vehicle $V_a$ and the associated RSU is long, i.e., the distance between vehicle and RSU is larger than or equal to 28.3 meters.

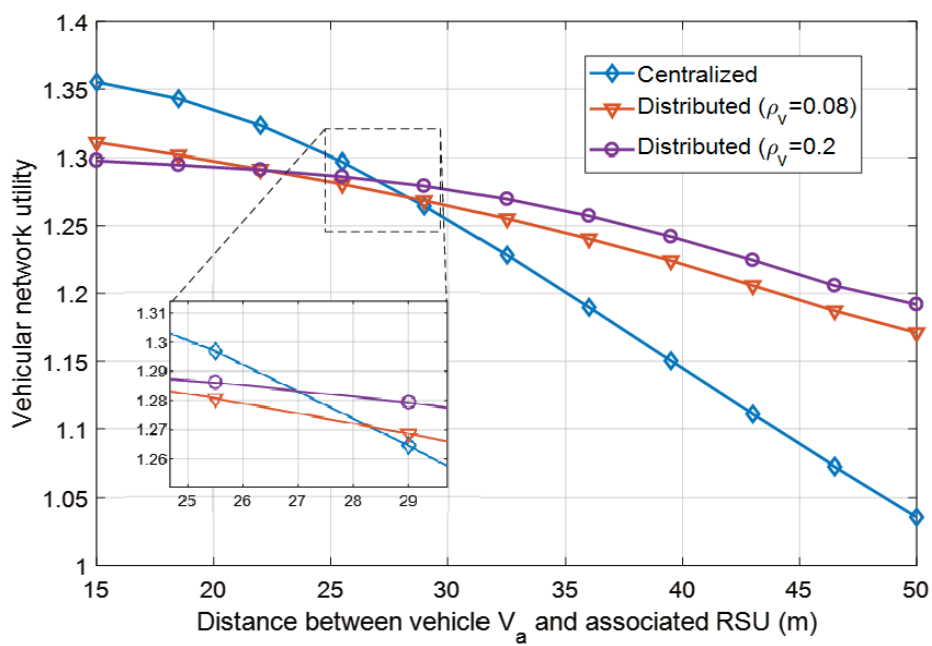

**Fig. 5**. Vehicular network utility with respect to the distance between vehicle $V_a$ and the associated RSU considering different vehicular network structures

In **Fig. 6(a)** and **Fig. 6(b)**, the impact of distance between vehicle $V_a$ and the associated RSU and network requirements on the vehicular network utility is investigated. When the network requirements are fixed, the utility of centralized vehicular network structure is higher than that of distributed vehicular network structure in the case that distance between vehicle $V_a$ and the associated RSU is less than a threshold. When the network requirement is more stringent, i.e., the solid lines indicates a stringent network requirement compared with dotted lines, the threshold will increase.

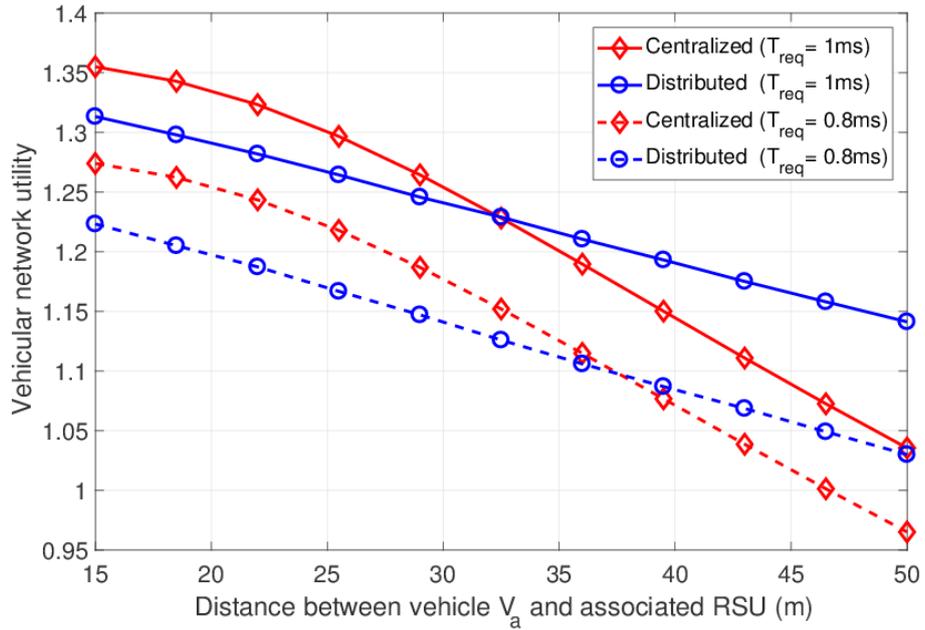

(a) Different latency requirements

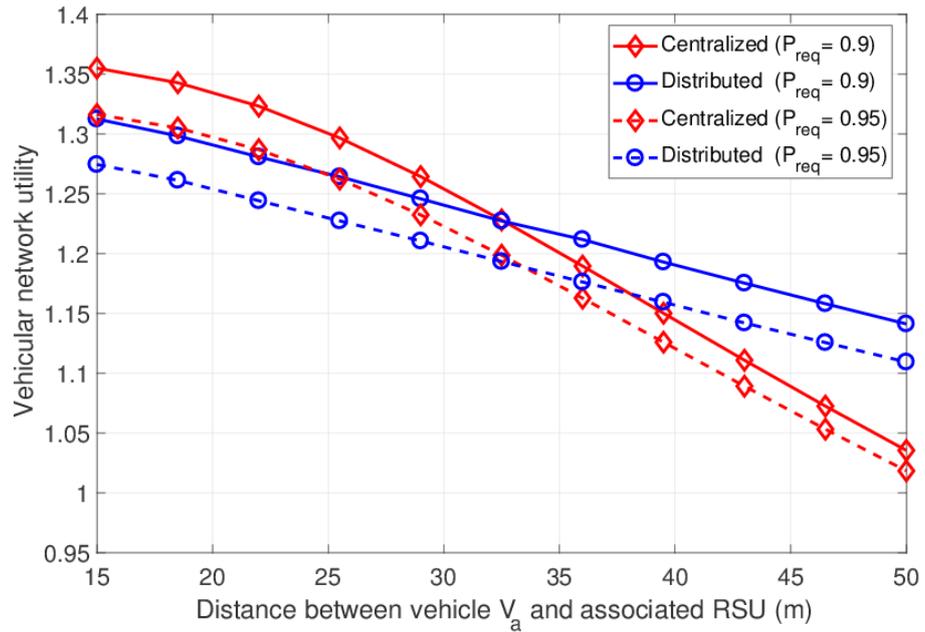

(b) Different reliability requirements

**Fig. 6**. Vehicular network utility with respect to the distance between vehicle $V_a$ and the associated RSU considering different network requirements

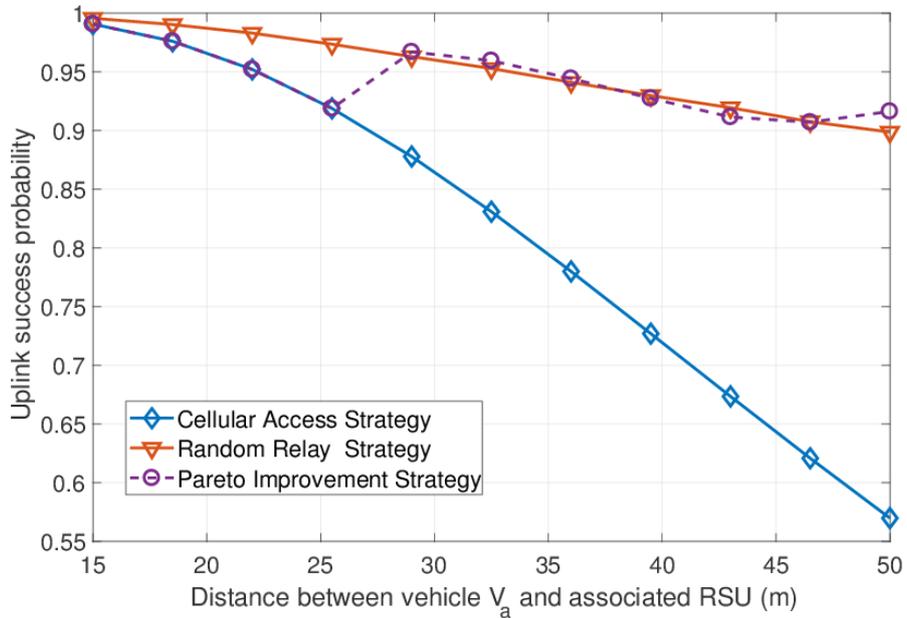

**Fig. 7**. Uplink success probability with respect to the distance between vehicle and the associated RSU considering different network access strategies

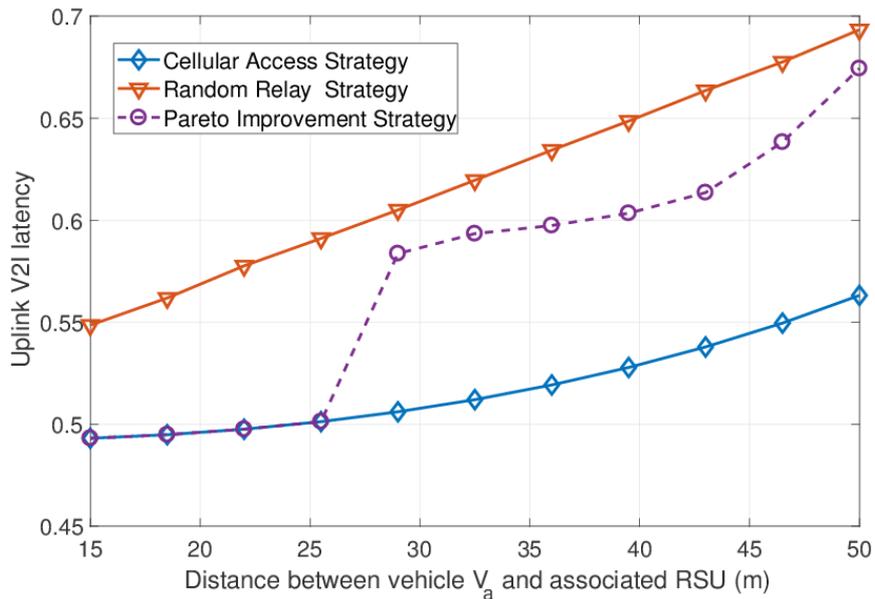

**Fig. 8**. Uplink V2I latency (ms) with respect to the distance between vehicle and the associated RSU considering different network access strategies

**Fig. 7** compares the Uplink success probability with respect to the distance between vehicle and the associated RSU considering three types of network access strategies. When the distance between vehicle and the associated RSU is less than or equal to 25 meters, the uplink success probability of Pareto improvement strategy is equal to the uplink success probability of cellular access strategy in vehicular networks. When the distance between vehicle and the associated RSU is larger than 25 meters, the uplink success probability of Pareto improvement strategy is close to the uplink success probability of random relay strategy.

**Fig. 8** compares the uplink V2I latency with respect to the distance between vehicle and the associated RSU considering three types of network access strategies. When the distance between vehicle and the associated RSU is less than 25 meters, the proposed Pareto improvement strategy has the same performance with the cellular access strategy. When the distance between vehicle and the associated RSU is larger than 25 meters, the uplink V2I latency of Pareto improvement strategy is less than the uplink V2I latency of random relay strategy and larger than the uplink V2I latency of cellular access strategy.

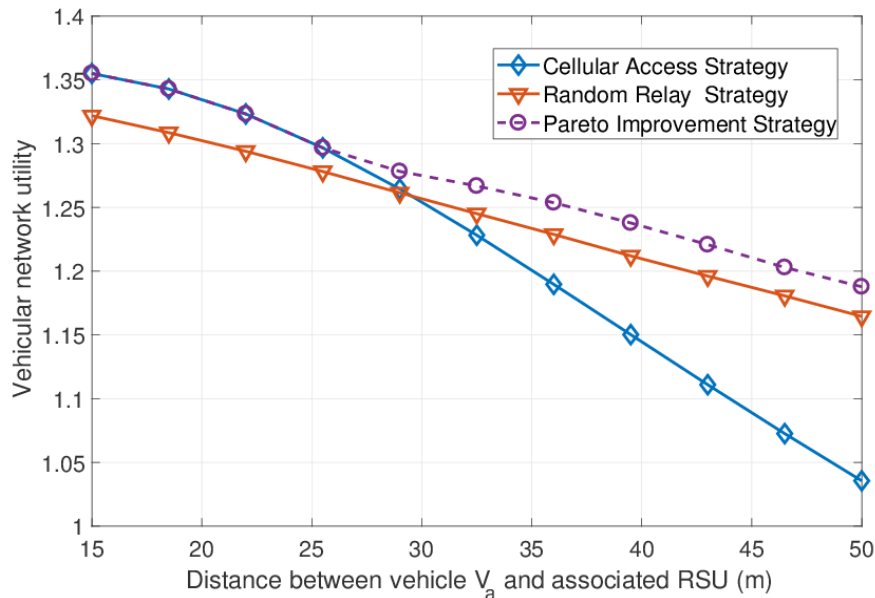

**Fig. 9**. Vehicular network utility with respect to the distance between vehicle and the associated RSU considering different network access strategies

**Fig. 9** shows the vehicular network utility comparison among three different network access strategies with respect to the distance between vehicle and the associated RSU. The cellular access strategy is the centralized network structure in vehicular networks. The random relay strategy means that each node chooses a network node randomly within its communication range as the relay node. The Pareto improvement strategy refers to the proposed uplink Pareto improvement algorithm is adopted in vehicular networks. When the network access strategy is fixed, the vehicular network utility decreases with increase of distance between vehicle and RSU. When the distance between vehicle and the associated RSU is larger than 25 meters, the vehicular network utility of Pareto improvement strategy always outperform the vehicular network utility of cellular access strategy and random relay strategy in vehicular networks. When the distance between vehicle and the associated RSU is less than or equal to 25 meters, the vehicular network utility of Pareto improvement strategy is equal to the vehicular network utility of cellular access strategy in vehicular networks. The Pareto improvement strategy is based on the Pareto improvement scheme which can lead to Pareto optimality on this problem, so this algorithm can always give an optimum utility function under different distances between vehicles and the associated RSU.

## 7. Conclusion

In this paper, a novel hybrid vehicular network framework combining centralized and distributed vehicular network structures was proposed. Based on this network framework, a new vehicular network utility model based on Euclidean norm theory was proposed to evaluate the latency and reliability utility of vehicular networks. Moreover, a vehicular network uplink optimization algorithm based on Pareto improvement was proposed to optimize the vehicular network utility. Simulation results indicate that the proposed uplink optimization algorithm can improve the uplink reliability and reduce the uplink latency£¬and the vehicular network utility is improved up to 15%.

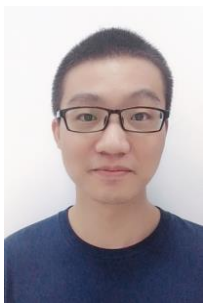

**Shikuan Li** received his Bachelor's degree in communication and information systems from Huazhong University of Science and Technology (HUST), Wuhan, China in in 2016, where he is currently working toward his Mas- ter's degree. His research interests include Hybrid vehicular network and 5G communication systems.

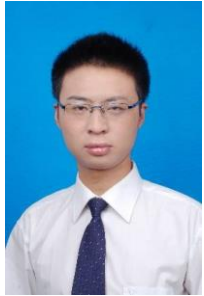

**Zipengli Li** received the B.E. degree in Telecommunication Engineering and M.S. degree in Communication and Information System from Huazhong University of Science and Technology (HUST), Wuhan, China in 2011 and 2014, respectively. He is currently working toward the Ph.D. degree in HUST. His research interests include vehicular networks and 5G mobile communication systems.

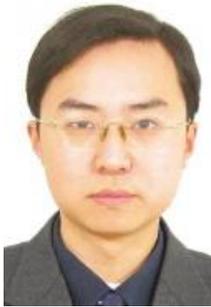

**Xiaohu Ge** is currently a full Professor with the School of Electronic Information and Communications at Huazhong University of Science and Technology (HUST), China. He is an adjunct professor with with the Faculty of Engineering and Information Technology at University of Technology Sydney (UTS), Australia. He received his PhD degree in Communication and Information Engineering from HUST in 2003. He has worked at HUST since Nov. 2005. Prior to that, he worked as a researcher at Ajou University (Korea) and Politecnico Di Torino (Italy) from Jan. 2004 to Oct. 2005. His research interests are in the area of mobile communications, traffic modeling in wireless networks, green communications, and interference modeling in wireless communications. He has published more than 200 papers in refereed journals and conference proceedings and has been granted about 25 patents in China. He received the Best Paper Awards from IEEE Globecom 2010. Dr. Ge served as the general Chair for the 2015 IEEE International Conference on Green Computing and Communications (IEEE GreenCom 2015). He serves as an associate editor for IEEE Wireless Communications, IEEE Transactions on Vehicular Technology and IEEE ACCESS, etc.

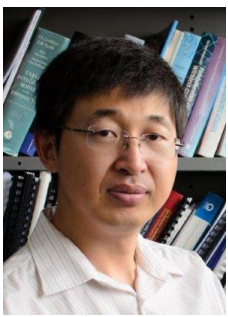

**Yonghui Li** received his PhD degree in November 2002 from Beijing University of Aeronautics and Astronautics. From 1999 – 2003, he was affiliated with Linkair Communication Inc, where he held a position of project manager with responsibility for the design of physical layer solutions for the LAS-CDMA system. Since 2003, he has been with the Centre of Excellence in Telecommunications, the University of Sydney, Australia. He is now a Professor in School of Electrical and Information Engineering, University of Sydney. He is the recipient of the Australian Queen Elizabeth II Fellowship in 2008 and the Australian Future Fellowship in 2012.

His current research interests are in the area of wireless communications, with a particular focus on MIMO, millimeter wave communications, machine to machine communications, coding techniques and cooperative communications. He holds a number of patents granted and pending in these fields. He is now an editor for IEEE transactions on communications and IEEE transactions on vehicular technology. He also served as a guest editor for several special issues of IEEE journals, such as IEEE JSAC special issue on Millimeter Wave Communications. He received the best paper awards from IEEE International Conference on Communications (ICC) 2014, IEEE PIMRC 2017 and IEEE Wireless Days Conferences (WD) 2014.